\documentclass[12pt]{article}
\newcommand{\xrm}[1]{{\textstyle \mbox{\rm #1}}}
\newcommand{\bm}[1]{\mbox{\boldmath $#1$}}
\def\chie{\mbox{\raisebox{0.5ex}{$\chi$}}}
\begin{document}
\title{\bf Scalar Mesons\\
within a model for all non-exotic mesons}
\author{{\large Eef van Beveren{\normalsize $^{\; a}$} and
George Rupp{\normalsize $^{\; b}$}}\\ [.5cm]
{\normalsize\it $^{a\;}$Centro de F\'{\i}sica Te\'{o}rica, Departamento de
F\'{\i}sica,}\\ {\normalsize\it Universidade, P3004-516 Coimbra, Portugal,}
{\small (eef@teor.fis.uc.pt)}\\ [.3cm]
{\normalsize\it $^{b\;}$Centro de F\'{\i}sica das Interac\c{c}\~{o}es
Fundamentais, Instituto Superior T\'{e}cnico,}\\ {\normalsize\it
Edif\'{\i}cio Ci\^{e}ncia,
P1049-001 Lisboa Codex, Portugal,} {\small (george@ajax.ist.utl.pt)}\\ [1cm]
{\normalsize Contribution to the}\\ {\normalsize\bf
Workshop on Recent Developments in Particle and Nuclear Physics}\\
{\normalsize at the}\\
{\normalsize Centro de F\'{\i}sica Te\'{o}rica da Universidade de Coimbra}
{\normalsize (30/04/2001)}
}
\maketitle

\begin{abstract}
We describe a four-parameter model for non-exotic meson-meson scattering,
which accommodates all non-exotic mesons, hence also the light scalar mesons,
as resonances and bound states characterised by complex singularities of the
scattering amplitude as a function of the total invariant mass.
The majority of the full $S$-matrix mesonic poles
stem from an underlying confinement spectrum.
However, the light scalar mesons $K_{0}^{\ast}(830)$, $a_{0}(980)$,
$f_{0}$(400--1200), and $f_{0}(980)$ do not,
but instead originate in $^{3}P_{0}$-barrier semi-bound states.

In the case of bound states, wave functions can be determined.
For $c\bar{c}$ and $b\bar{b}$, radiative transitions have been calculated.
Here we compare the results to the data.
\end{abstract}
\clearpage

\section{Introduction}

Strong interactions have been under study for over half a century by now.
Starting from curiosity over nuclear forces \cite{Yukawa35},
they have been developing towards the need to understand
Quantum Chromodynamics (QCD) \cite{SchQCD01}.
The modelling and analysis of meson-meson scattering is thereby believed
to give valuable contributions.
Here we will report on a model for meson-meson scattering which has been
very successful in predicting and reproducing a host of experimental data
with a very limited number of parameters
\cite{BR01b}--\cite{BDR80}.
It is based on our perspective of QCD, as outlined below.

Mesons are composed of quarks, antiquarks, and gluons,
whether stable with respect to OZI-allowed hadronic decay,
i.e., bound states like $K$, $J/\psi$, and $\Upsilon$,
or unstable, decaying into two or more lighter mesons,
like e.g.\ the resonances $\rho$, $\omega$, and $\phi$.
Mesons can be described by systems of one constituent quark ($q$)
confined to one constituent antiquark ($\bar{q}$).
Constituent quarks and antiquarks, which can be thought of as lumps of bare
quarks surrounded by glue and a cloud of virtual $q\bar{q}$ pairs,
determine the flavour of the meson and the bulk of its mass.
The remaining strong interactions may be parametrised by an effective
confining potential, e.g.\ a harmonic oscillator.

Decay of a meson into a pair of lighter mesons is assumed to be the result
of the creation of a new constituent $q\bar{q}$ pair with the quantum
numbers of the vacuum ($^{3}P_{0}$), leading to OZI-allowed transitions.
Mesonic resonances in meson-meson scattering are supposed to be formed through
the inverse phenomenon,
i.e., the annihilation of a constituent $q\bar{q}$ pair
with the quantum numbers of the vacuum, one from each of the two mesons
involved in the scattering process.

The spatial quantum numbers ($\nu ,J^{PC}\,$) of a meson,
either bound state or resonance,
follow from the total spin of the constituent $q\bar{q}$ pair,
their relative orbital angular momentum $\ell$,
and their radial quantum number $\nu$.

\begin{figure}[ht]
\begin{center}
\begin{picture}(393,225)(-138,27)
\put(5,125){\makebox(0,0)[rc]{$\left.\begin{array}{c}
\xrm{confinement}\\ +\\
^{3}P_{0}\;\; q\bar{q}\;\;\xrm{creation}\end{array}
\right\}\;\;$\Large\bm{\Longrightarrow}}}
\put(115,125){\makebox(0,0)[lc]{{\Large\bm{\Longrightarrow}}
$\left\{\begin{array}{c} \xrm{Meson + Meson}\\
\xrm{scattering}\\ \xrm{cross sections}\\
\xrm{resonances}\\
\xrm{bound states}\end{array}\right.$}}
\put(-60,75){\makebox(0,0)[tl]{
$\left.\begin{array}{c} J^{PC}\\
\xrm{isospin}\\ \xrm{flavours}\end{array}\right\}
\begin{array}{c} \xrm{quantum numbers}\\
\xrm{(variables)}\end{array}$}}
\put(-140,160){\makebox(0,0)[bl]{
$\left.\begin{array}{c} \omega \xrm{(confinement)}\\
V_{t}\left\{\begin{array}{l}
\lambda \xrm{(intensity)}\\
a \xrm{(form factor)}\\
g_{c} \xrm{(colour splitting)}\end{array}\right.\\ \\ [-0.4cm]
\xrm{effective quark masses}\\
(n,s,c,b)
\end{array}\right\}$
parameters}}
\put(60,125){\makebox(0,0){\Large\bf MODEL}}
\end{picture}
\end{center}
\caption[]{Schematic picture of the model for the description of
non-exotic meson-meson scattering.}
\label{MODEL}
\end{figure}
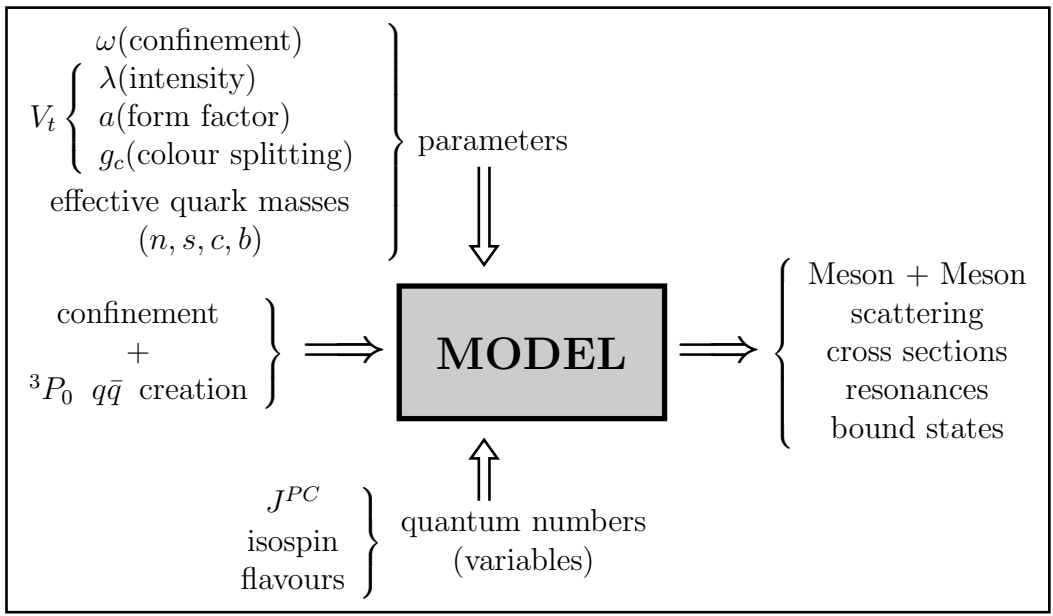

In Fig.~(\ref{MODEL}) we schematically show how the model for the
description of non-exotic meson-meson scattering works.
The employed dynamics is characterised by confinement and $^{3}P_{0}$
$q\bar{q}$-pair creation.
Confinement is provided by a harmonic oscillator potential with universal
frequency $\omega$.
On the other hand, $^{3}P_{0}$ $q\bar{q}$-pair creation is modelled
by a transition potential $V_{t}$, which has two parameters,
that is, the universal overall transition intensity $\lambda$,
and a distance parameter $a$
that is related to the shape of the transition potential.
The so-called colour splitting of pseudo-scalar and vector mesons is
absorbed in the transition potential through the parameter $g_{c}$, at least
in the present version of the model.

The four model parameters and the four constituent quark masses for
non-strange (either $u$ or $d$), strange, charm, and bottom quarks
are adjusted to the data, i.e., to partial-wave cross sections of meson-meson
scattering, or to mesonic bound-state and resonance spectra.
Model predictions can then be obtained by giving as input data
the relevant quantum numbers, namely flavour, isospin, total angular momentum,
parity, and possibly charge conjugation.
The outputs of the model are not only predictions for scattering quantities,
but also comprise bound-state spectra and corresponding wave functions for the
mesonic systems under study.

The philosophy of the model is simple:
in the scattering region for a meson-meson scattering process
we assume the coexistence of a variety of different systems or channels, that
is, one or more confined $q\bar{q}$ pairs, and
all combinations of meson-meson pairs that have the correct quantum numbers.
For instance, when studying scattering in an isosinglet, we assume
that there are definitely non-zero probabilities to find in the interaction
region an $n\bar{n}$ pair, an $s\bar{s}$ pair, but also a $\pi\pi$ pair, a
$K\bar{K}$ pair, and so forth.
The first type of channels differs from the second type in that
$q\bar{q}$ pairs are permanently confined, whereas meson-meson pairs
feel more modest forces, which might even be neglected in lowest order.

When we describe quarkonia by wave functions $\psi_{c}$ and two-meson
systems by wave functions $\psi_{f}$, then we obtain for their time evolution
the wave equations

\begin{equation}
\left( E-H_{c}\right)\;\psi_{c}\left(\vec{r}\;\right)\; =\;
V_{t}\;\psi_{f}\left(\vec{r}\;\right)
\;\;\;\xrm{and}\;\;\;
\left( E-H_{f}\right)\;\psi_{f}\left(\vec{r}\;\right)\; =\;
\left[ V_{t}\right]^{T}\;\psi_{c}\left(\vec{r}\;\right)
\;\;\; .
\label{cpldeqna}
\end{equation}

\noindent
Here, $H_{c}$ describes the dynamics of confinement in the interaction region,
$H_{f}$ the dynamics of the scattered particles, and $V_{t}$ the communication
between the two different sectors.

For the dynamics of confinement we understand here that, as a function of
the interquark distance $r$, the resulting quark-antiquark binding forces
grow rapidly outside the interaction region.
Consequently, we must eliminate $\psi_{c}$ from the equations (\ref{cpldeqna}),
since it is vanishing at large distances and thus {\it unobservable}.
Formally, it is easy to do so. We thus obtain the relation

\begin{equation}
\left( E-H_{f}\right)\;\psi_{f}\left(\vec{r}\;\right)\; =\;
\left[ V_{t}\right]^{T}\;
\left( E-H_{c}\right)^{-1}\; V_{t}\;\psi_{f}\left(\vec{r}\;\right)
\;\;\; .
\label{scatteqna}
\end{equation}

The set of coupled equations (\ref{scatteqna}) for the two-meson channels
represented by $\psi_{f}$ can be solved for the scattering amplitude,
either numerically or analytically,
once $H_{c}$, $H_{f}$, and $V_{t}$ are specified.
\clearpage

\section{Fixing the parameters}

The two parameters $\lambda$ and $a$ of the transition potential $V_{t}$
can be fixed, together with the charm and bottom masses,
by the spectra of the $c\bar{c}$ and $b\bar{b}$ vector states.
A comparison of model results with experiment is shown in Table
(\ref{ccbb}).
In the first column of the table we find the spectroscopic notation
for the various states:
the first symbol indicates the quarkonium radial excitation $N=\nu +1$,
which for the here employed harmonic-oscillator confinement
is related to the total bare quarkonium mass $E_{\ell\nu}$ by

\begin{equation}
E_{\ell\nu}\; =\;\omega\left( 2\nu +\ell +\frac{3}{2}\right)\;
+\;\xrm{quark masses}
\;\;\; .
\label{Eellnu}
\end{equation}
\begin{table}[ht]
\begin{center}
\begin{tabular}{|c||rr|rr|l|c|}
\hline\hline & & & & & \multicolumn{2}{c|}{ } \\ [-0.3cm]
 & \multicolumn{2}{c|}{$c\bar{c}$}
& \multicolumn{2}{c|}{$b\bar{b}$} & \multicolumn{2}{c|}{ } \\
$N^{(2s+1)}\ell_{J}$ & model & exp. & model & exp. &
\multicolumn{2}{c|}{meson-meson channels involved}\\
\hline & & & & & & \\ [-0.3cm]
 & GeV & GeV & GeV & GeV & & $(L,S)$ \\ [.2cm]
1$^{3}$S$_{1}$ &  3.10 & 3.10 & 9.41 & 9.46 & $c\bar{c}$ sector & \\
2$^{3}$S$_{1}$ &  3.67 & 3.69 & 10.00 & 10.02 &
$DD$, $D_{s}D_{s}$, & (1,0)\\
1$^{3}$D$_{1}$ &  3.80 & 3.77 & 10.14 & ... &
$DD^{\ast}$, $D_{s}D_{s}^{\ast}$, & (1,1)\\
3$^{3}$S$_{1}$ &  4.05 & 4.04 & 10.40 & 10.36 &
$D^{\ast}D^{\ast}$, $D_{s}^{\ast}D_{s}^{\ast}$ & (1,0), (1,2) and (3,2)\\
\cline{6-7}
2$^{3}$D$_{1}$ &  4.14 & 4.16 & 10.48 & ...  &
\raisebox{-0.5mm}{$b\bar{b}$ sector} & \\
4$^{3}$S$_{1}$ &  4.41 & 4.42 & 10.77 & 10.58 &
$BB$, $B_{s}B_{s}$, & (1,0)\\
3$^{3}$D$_{1}$ &  ...  & ...  & 10.86 & 10.87 &
$BB^{\ast}$, $B_{s}B_{s}^{\ast}$, & (1,1)\\
5$^{3}$S$_{1}$ &  ...  & ...  & 11.15 & 11.02 &
$B^{\ast}B^{\ast}$, $B_{s}^{\ast}B_{s}^{\ast}$ & (1,0), (1,2) and (3,2)\\
\hline\hline
\end{tabular}
\end{center}
\caption[]{Comparison of the model results \cite{BRRD83,BDR80} for
$J^{PC}=1^{--}$ charmonium and bottomonium bound-state masses and resonance
central mass positions to experiment \cite{PDG}.}
\label{ccbb}
\end{table}
In the last column of Table (\ref{ccbb}) we find the meson-meson channels
which are involved in the determination of the quarkonia spectra.
Vector-vector channels come in pairs, one for total spin zero
and one for total spin equal to two. Moreover, spin 2 may combine with $F$
waves.
The complete wave function of charmonium (bottomonium) is thus composed of
twelve channels, the two $c\bar{c}$ ($b\bar{b}$) channels,
one for $S$ wave and one for $D$ wave, coupled to ten meson-meson channels
through $^{3}P_{0}$ transitions. Note that not all of these channels are open
for the systems under consideration, but that does not necessarily mean their
influence is negligible.
The relative coupling intensities for the various channels,
which follow from the three-meson vertices defined in
Refs. \cite{Bev83,Bev84}, are tabulated in Ref.~\cite{Zweig}. In Nature,
many more meson-meson channels couple in principle, but in the present
version of the model only pseudoscalar and vector mesons are considered
in the initial and final scattering states, since these are generally the ones
that lie close enough to have an appreciable effect.

\begin{figure}[ht]
\begin{center}
\begin{picture}(170.08,160.08)(-35.50,-21.30)
\put(15.89,-3.92){\makebox(0,0)[tc]{0.6}}
\put(55.62,-3.92){\makebox(0,0)[tc]{0.8}}
\put(95.35,-3.92){\makebox(0,0)[tc]{1.0}}
\put(-3.92,41.48){\makebox(0,0)[rc]{50}}
\put(-3.92,82.96){\makebox(0,0)[rc]{100}}
\put(-3.92,124.45){\makebox(0,0)[rc]{150}}
\put(135.08,-3.92){\makebox(0,0)[tr]{GeV}}
\put(63.53,-14.25){\makebox(0,0)[tc]{$\pi\pi$ invariant mass}}
\put(-3.92,145.32){\makebox(0,0)[tr]{phase}}
\put(3.92,141.31){\makebox(0,0)[tl]{$\pi\pi$ $P$-wave}}
\put(90,55){\makebox(0,0){\huge\bf\bm{m_{n}}}}
\put(5.96,8.30){\makebox(0,0){$\bullet$}}
\put(20.86,15.76){\makebox(0,0){$\bullet$}}
\put(28.80,24.89){\makebox(0,0){$\bullet$}}
\put(33.77,32.36){\makebox(0,0){$\bullet$}}
\put(37.74,39.82){\makebox(0,0){$\bullet$}}
\put(41.72,49.78){\makebox(0,0){$\bullet$}}
\put(44.70,58.08){\makebox(0,0){$\bullet$}}
\put(46.68,63.88){\makebox(0,0){$\bullet$}}
\put(48.67,70.52){\makebox(0,0){$\bullet$}}
\put(50.66,76.33){\makebox(0,0){$\bullet$}}
\put(52.64,82.14){\makebox(0,0){$\bullet$}}
\put(54.63,87.11){\makebox(0,0){$\bullet$}}
\put(57.61,94.58){\makebox(0,0){$\bullet$}}
\put(61.58,102.88){\makebox(0,0){$\bullet$}}
\put(65.56,108.68){\makebox(0,0){$\bullet$}}
\put(69.53,112.83){\makebox(0,0){$\bullet$}}
\put(73.50,117.81){\makebox(0,0){$\bullet$}}
\put(77.47,120.30){\makebox(0,0){$\bullet$}}
\put(82.44,124.45){\makebox(0,0){$\bullet$}}
\put(88.40,126.94){\makebox(0,0){$\bullet$}}
\put(95.35,129.43){\makebox(0,0){$\bullet$}}
\put(103.30,131.08){\makebox(0,0){$\bullet$}}
\put(110.25,132.74){\makebox(0,0){$\bullet$}}
\put(116.21,134.40){\makebox(0,0){$\bullet$}}
\put(122.17,135.23){\makebox(0,0){$\bullet$}}
\put(125.15,136.06){\makebox(0,0){$\bullet$}}
\end{picture}
\begin{picture}(190.08,160.08)(-55.50,-21.30)
\put(0.00,-3.92){\makebox(0,0)[tc]{0.7}}
\put(27.02,-3.92){\makebox(0,0)[tc]{0.8}}
\put(54.03,-3.92){\makebox(0,0)[tc]{0.9}}
\put(81.05,-3.92){\makebox(0,0)[tc]{1.0}}
\put(-3.92,41.48){\makebox(0,0)[rc]{50}}
\put(-3.92,82.96){\makebox(0,0)[rc]{100}}
\put(-3.92,124.45){\makebox(0,0)[rc]{150}}
\put(135.08,-3.92){\makebox(0,0)[tr]{GeV}}
\put(63.53,-14.25){\makebox(0,0)[tc]{$K\pi$ invariant mass}}
\put(-3.92,145.32){\makebox(0,0)[tr]{phase}}
\put(3.92,141.31){\makebox(0,0)[tl]{$K\pi$ $P$-wave}}
\put(90,55){\makebox(0,0){\huge\bf\bm{m_{s}}}}
\put(8.11,4.98){\makebox(0,0){$\bullet$}}
\put(21.61,8.30){\makebox(0,0){$\bullet$}}
\put(32.42,11.62){\makebox(0,0){$\bullet$}}
\put(39.17,19.83){\makebox(0,0){$\bullet$}}
\put(41.88,23.48){\makebox(0,0){$\bullet$}}
\put(44.58,32.77){\makebox(0,0){$\bullet$}}
\put(47.28,41.15){\makebox(0,0){$\bullet$}}
\put(49.98,58.57){\makebox(0,0){$\bullet$}}
\put(52.68,79.81){\makebox(0,0){$\bullet$}}
\put(55.38,89.85){\makebox(0,0){$\bullet$}}
\put(58.09,101.55){\makebox(0,0){$\bullet$}}
\put(60.79,111.67){\makebox(0,0){$\bullet$}}
\put(63.49,116.48){\makebox(0,0){$\bullet$}}
\put(66.19,122.29){\makebox(0,0){$\bullet$}}
\put(68.89,125.19){\makebox(0,0){$\bullet$}}
\put(75.65,130.25){\makebox(0,0){$\bullet$}}
\put(86.45,135.15){\makebox(0,0){$\bullet$}}
\put(97.26,137.06){\makebox(0,0){$\bullet$}}
\put(108.07,138.97){\makebox(0,0){$\bullet$}}
\put(118.87,139.80){\makebox(0,0){$\bullet$}}
\put(129.68,143.53){\makebox(0,0){$\bullet$}}
\end{picture}
\end{center}
\caption[]{Comparison of the model results (solid lines \cite{BRRD83}) for the
$\pi\pi$ and $K\pi$ elastic scattering phase shifts with $J^{P}=1^{-}$
quantum numbers to experiment ($\bullet$), in the energy regions of the
$\rho (770)$ and $K^{\ast}(892)$ mesons, respectively.
The data for $\pi\pi$ are taken from Ref.~\cite{Proto73},
and for $K\pi$ from Ref.~\cite{ECMDLL78}.}
\label{ns}
\end{figure}
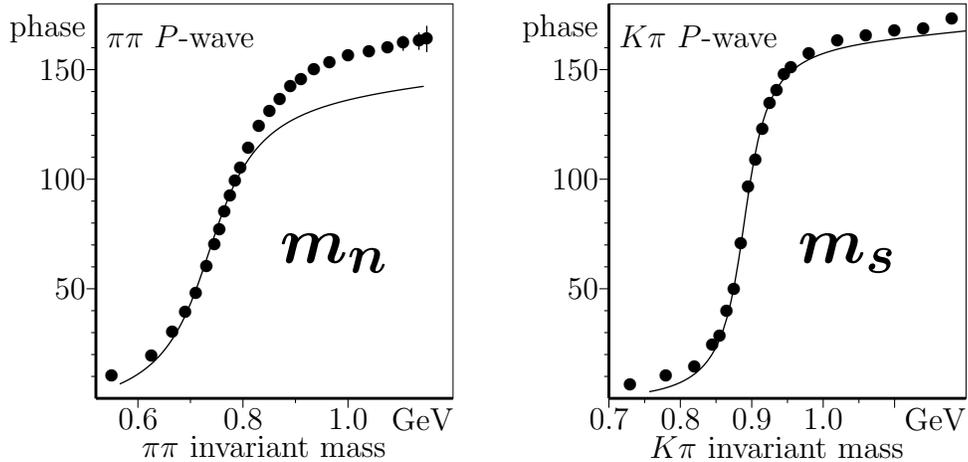

The non-strange and strange constituent quark masses are 
fixed by the elastic $P$-wave $\pi\pi$ and $K\pi$ scattering phase shifts,
respectively, as shown in Fig.~(\ref{ns}).
We observe from the first plot that the agreement between model and experiment
is not perfect for larger values of the $\pi\pi$ invariant mass.
We believe this to be a consequence of neglecting residual strong forces
in meson-meson scattering that are not due to the $^{3}P_{0}$ mechanism,
like meson exchange.
This might also explain why for $K\pi$ scattering the model gives superior
results.
Nevertheless, it is remarkable that a model which describes the $c\bar{c}$ and
$b\bar{b}$ quarkonia bound states and resonances very well, can predict to a
fair precision the $\pi\pi$ and $K\pi$ scattering data, just by fixing
reasonable values for $m_{n}$ and $m_{s}$.  Furthermore, we notice that the
correct slopes of the phase shifts at $90^{\circ}$, as a function of the
invariant masses, imply the model predicts the right widths for the $\rho$(770)
and the $K^{\ast}(892)$.

Finally, the parameter $g_{c}$, which accounts for the hyperfine mass splitting
of pseudoscalar and vector mesons, is fixed by the mass differences between the
$\pi\eta\eta 'K$ and $\rho\omega\phi K^{\ast}$ nonets.
In Table~(\ref{mass}) we summarise the model results \cite{BRRD83}
for the light vector-meson sector.
\begin{table}[ht]
\begin{center}
\begin{tabular}{|c||llc|l|c|}
\hline\hline & & & & \multicolumn{2}{c|}{ } \\ [-0.3cm]
state & model & exp. & Ref. &
\multicolumn{2}{c|}{meson-meson channels involved}\\
\hline & & & & & \\ [-0.3cm]
 & GeV & GeV & & & $(L,S)$ \\
$\rho (1S)$ &  0.76 & 0.77 & \cite{PDG} & & \\
$\rho (2S)$ & 1.29 & 1.29 & \cite{LASS94} &
$\pi\pi$, $KK$, & (1,0)\\
$\rho (1D)$ & 1.40 & 1.42 & \cite{DFM88} &
$\pi\omega$, $\eta\rho$, $KK^{\ast}$, & (1,1)\\
$\rho (3S)$ & 1.59 & 1.60 & \cite{ChDu91} &
$\rho\rho$, $K^{\ast}K^{\ast}$ & (1,0), (1,2) and (3,2)\\
$\rho (2D)$ & 1.68 & 1.68 & \cite{DFM88} & & \\
\hline & & & & & \\
$K^{\ast}(1S)$ & 0.93 & 0.89 & \cite{PDG} &
$\pi K$, $\eta K$, $\eta 'K$, & (1,0)\\
$K^{\ast}(2S)$ & 1.41 & 1.41 & \cite{PDG} &
$\pi K^{\ast}$, $\eta K^{\ast}$, $\eta 'K^{\ast}$, $K\rho$,
$K\omega$, $K\phi$, & (1,1)\\
$K^{\ast}(3S)$ & 1.73 & 1.72 & \cite{PDG} &
$\rho K^{\ast}$, $\omega K^{\ast}$, $\phi K^{\ast}$ & (1,0), (1,2) and (3,2)\\
\hline & & & & & \\
$\phi (1S)$ & 1.03 & 1.02 & \cite{PDG} & $KK$, & (1,0)\\
$\phi (2S)$ & 1.53 & - & & $\eta '\phi$, $KK^{\ast}$, & (1,1)\\
$\phi (3S)$ & 1.87 & - & & $K^{\ast}K^{\ast}$ & (1,0), (1,2) and (3,2)\\
\hline\hline
\end{tabular}
\end{center}
\caption[]{Real parts of the singularities in the meson-meson scattering
matrices for some of the well-known light vector resonances.}
\label{mass}
\end{table}
We find good agreement between model and experiment.
However, with respect to the radial excitations of the $\rho (770)$ meson,
we believe that the particles refered to in Ref.~\cite{PDG},
are the $D$ states.
In the harmonic-oscillator spectrum a $D$ state is degenerate with
the $S$ state belonging to the next radial excitation.
Through the coupling to the meson-meson channels the degeneracy is lifted,
which gives rise to separate resonances (see Fig.~\ref{charm} below).
Also the $\phi (1680)$ corresponds in our model to the $\phi (1D)$ state.

For the pseudoscalars and the open-charm states, not shown here,
the model predictions \cite{BRRD83} manifest much more disagreement with the
data. Possible causes are a deficient treatment of chiral symmetry, relativity,
and pseudothresholds. Further study on these points is clearly required, which
is already in course.
\clearpage

\section{Resonances}

The overall transition parameter $\lambda$ provides the communication
between permanently closed $q\bar{q}$ channels and meson-meson
scattering channels.
To lowest order, this describes the decay of a meson, but to higher order also
the occurrence of resonances in meson-meson scattering.
For $\lambda=0$, the set of equations (\ref{cpldeqna}) describes free,
non-interacting two-meson systems with a continuum spectrum above threshold,
and confined $q\bar{q}$ systems with a discrete mass spectrum,
the so-called {\it bare} \/mesons, as given by formula (\ref{Eellnu}) for
non-negative integer values of the radial quantum number $\nu$.
When $\lambda$ is non-zero, then, for energies above threshold,
the model describes the scattering of interacting two-meson systems,
elastic as well as non-elastic,
whereas, for energies below threshold, the model describes physical mesons
that are stable with respect to OZI decays.
A resonance does not represent just one single state, but rather a continuum of
states.  Each state under the resonance has a different mass, whereby the
intensities of its couplings to the different two-meson systems are
proportional to the corresponding partial-wave scattering cross sections at
that energy. Hence, a resonance should be represented by a whole range of
masses, though with varying importance for the scattering data.
Such a phenomenon can also be described by a single complex number,
namely a singularity in the analytically continued scattering amplitude.
For narrow resonances, which here correspond to small values of $\lambda$,
the real and imaginary parts roughly equal the central resonance mass
and half its width, respectively.
However, for broad and highly deformed resonances, corresponding to the
realm of strong interactions in our model, such simple relations do not exist.
It would be helpful if in experiment complex masses could be measured,
but exciting the imaginary parts of masses at particle accellerators goes far
beyond our imagination as yet.
At best, experiment supplies us with partial-wave cross sections, for which
scattering amplitudes have to be constructed.
Breit-Wigner shapes \cite{BW} are the usual techniques, though
for strong interactions probably not the most adequate tools.
Moreover, since each scattering amplitude has a different pole structure, no
generally applicable method exists how to locate singularities.
Accordingly, since these poles, with the respective residues, constitute the
most relevant information on scattering properties, it would be helpful if the
Particle Data Group Collaboration
could show us the world averages of full shapes
of elastic and inelastic partial-wave cross sections or phase shifts,
rather then just the central masses and widths of resonances.

For the time being, it seems that we must content ourselves with at least
agreement on
the number of scattering-amplitude singularities within large intervals of the
invariant mass. For small values of $\lambda$, complex poles in the relevant
Riemann sheet are expected to lie close to the bare meson masses, whereas for
increasing $\lambda$ the singularities move farther away.  Hence, one might
expect that their number agrees with the number of bare mesons.  Consequently,
bound states and resonances could then both be labelled by the spectroscopic
quantum numbers, as for instance shown in Table~(\ref{ccbb}). Nevertheless,
in the case of the scalar mesons we will see below that not all singularities
stem from the bare spectrum.

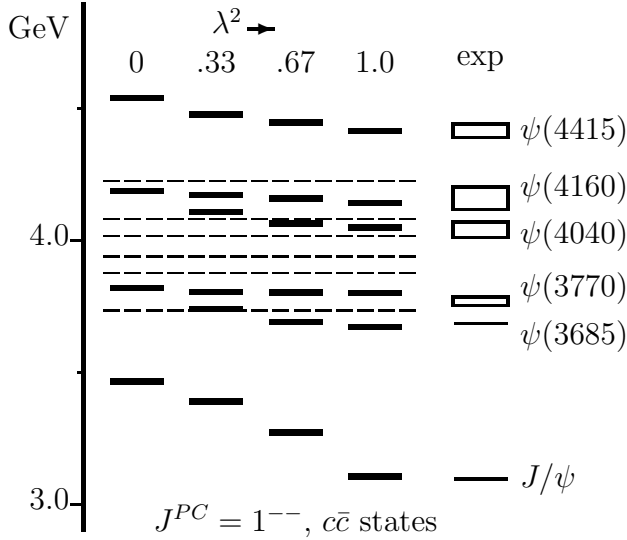
\begin{figure}[ht]
\begin{picture}(250,220)(-50,270)
\put(80,300){\makebox(0,0)[t]{$J^{PC}=1^{--}$, $c\bar{c}$ states}}
\linethickness{1.5pt}
\put(0,290){\line(0,1){200}}
\thicklines
\put(-5.0,300){\line(1,0){5}}
\put(-5.5,306){\makebox(0,0)[tr]{3.0}}
\put(-5.0,400){\line(1,0){5}}
\put(-5.5,406){\makebox(0,0)[tr]{4.0}}
\put(-2.5,350){\line(1,0){2.5}}
\put(-2.5,450){\line(1,0){2.5}}
\put(-5.5,485){\makebox(0,0)[tr]{GeV}}
\put(60,488){\makebox(0,0)[tr]{$\lambda^{2}$}}
\put(62,480){\vector(1,0){10}}
\put(20,464){\makebox(0,0)[b]{0}}
\put(50,464){\makebox(0,0)[b]{.33}}
\put(80,464){\makebox(0,0)[b]{.67}}
\put(110,464){\makebox(0,0)[b]{1.0}}
\put(150,464){\makebox(0,0)[b]{exp}}
\linethickness{2.0pt}
\put(10,454.0){\line(1,0){20}}
\put(10,418.8){\line(1,0){20}}
\put(10,382.0){\line(1,0){20}}
\put(10,346.5){\line(1,0){20}}
\put(40,447.7){\line(1,0){20}}
\put(40,417.1){\line(1,0){20}}
\put(40,410.9){\line(1,0){20}}
\put(40,380.5){\line(1,0){20}}
\put(40,374.1){\line(1,0){20}}
\put(40,339.0){\line(1,0){20}}
\put(70,444.7){\line(1,0){20}}
\put(70,415.9){\line(1,0){20}}
\put(70,406.4){\line(1,0){20}}
\put(70,380.3){\line(1,0){20}}
\put(70,369.2){\line(1,0){20}}
\put(70,327.3){\line(1,0){20}}
\put(100,441.3){\line(1,0){20}}
\put(100,414.3){\line(1,0){20}}
\put(100,404.8){\line(1,0){20}}
\put(100,380.1){\line(1,0){20}}
\put(100,367.3){\line(1,0){20}}
\put(100,310.7){\line(1,0){20}}
\thicklines
\put(140,439.4){\framebox(20,4.3){}}
\put(165,441.5){\makebox(0,0)[lc]{$\psi$(4415)}}
\put(140,412.1){\framebox(20,7.8){}}
\put(165,419.9){\makebox(0,0)[lc]{$\psi$(4160)}}
\put(140,401.4){\framebox(20,5.2){}}
\put(165,401.9){\makebox(0,0)[lc]{$\psi$(4040)}}
\put(140,375.8){\framebox(20,2.4){}}
\put(165,382.0){\makebox(0,0)[lc]{$\psi$(3770)}}
\put(140,368.6){\line(1,0){20}}
\put(165,364.0){\makebox(0,0)[lc]{$\psi$(3685)}}
\put(140,309.7){\line(1,0){20}}
\put(165,309.7){\makebox(0,0)[lc]{$J/\psi$}}
\thinlines
\multiput(7.5,422.5)(8,0){15}{\line(1,0){6}}
\multiput(7.5,408.2)(8,0){15}{\line(1,0){6}}
\multiput(7.5,393.9)(8,0){15}{\line(1,0){6}}
\multiput(7.5,401.7)(8,0){15}{\line(1,0){6}}
\multiput(7.5,387.6)(8,0){15}{\line(1,0){6}}
\multiput(7.5,373.4)(8,0){15}{\line(1,0){6}}
\end{picture}
\caption[]{The model results \cite{BRRD83,BDR80} for the central resonance
positions of charmonium $^3S_1$ and $^3D_1$ states, 
for four values of the parameter $\lambda$,
compared to the experimental situation \cite{PDG}.
The various dashed lines indicate the threshold positions of the strong decay
channels $DD$, $DD^{\ast}$, $D_{s}D_{s}$, $D^{\ast}D^{\ast}$,
$D_{s}D_{s}^{\ast}$, and $D_{s}^{\ast}D_{s}^{\ast}$.}
\label{charm}
\end{figure}

Bare masses represent bare mesons that result from permanent confinement only.
Therefore, models restricted to this particular sector of strong interactions
should not adjust their parameters to the bound states and resonances from
experiment, although this is common practice unfortunately.
In Fig.~(\ref{charm}) we depict how the real parts of the poles for the vector
charmonium states behave in our model, when $\lambda$ is varied.
For $\lambda =0$, the bare spectrum of formula (\ref{Eellnu}) is shown.
For $\lambda =1$, one obtains the model result of Table~(\ref{ccbb}).
Note that the effect of open-charm channels on the ground state is highly
non-perturbative, despite of these channels being all virtual.

Bound states can be represented by special singularities of the
scattering amplitude, i.e., poles on the real energy axis
below the lowest threshold of the system.
The $J/\psi$ (1S) is an example of such a state (see Fig.~\ref{charm}).
It consists of two $c\bar{c}$ channels, a dominant one
in $S$ wave, the other one in $D$ wave, as well as of channels with virtual
pairs of $D$, $D_{s}$, $D^{\ast}$, and $D_{s}^{\ast}$ mesons in $P$ or $F$
waves. The $\psi$(2S) is an example of a bound state which originates in a
bare meson located in the open-charm continuum. For values of $\lambda$ up to
$0.6$, the $\psi$(2S) singularity of the scattering amplitude has an imaginary
part, which corresponds to a resonance. But at the $D\bar{D}$ threshold, the
$\psi$(2S) pole turns real, for $\lambda \approx 0.6$. For the model value of
$\lambda =1$, the singularity is found on the real axis below the $D\bar{D}$
threshold, in accordance with the physical $\psi$(2S) state.
\clearpage

\section{Scalar mesons}

In the foregoing, we have discussed how the parameter $\lambda$,
which describes the overall intensity of the coupling of
$q\bar{q}$ systems to meson-meson scattering channels through 
the $^{3}P_{0}$ mechanism, can be adjusted to the data.
When we carefully inspect wave equation (\ref{scatteqna}),
then we observe that the centrifugal barrier is the main contribution
of the left-hand side of the equation,
whereas the right-hand side is proportional to $\lambda^{2}$.
Consequently, for $S$-wave scattering, in the absence of a centrifugal
barrier, the right-hand side of Eq.~(\ref{scatteqna}) dominates
the interaction.
Now, for low energies far away from the bare states, this term has the form
of a potential barrier, in which loosely bound two-meson systems may exist,
solely depending on the value of $\lambda$.
This is exactly what happens for the light flavours.
The $^{3}P_{0}$ barrier, not high enough to form bound states, gives rise
to a complete nonet of light scalar resonances with central mass positions at
about 0.8--1.0 GeV.

\begin{figure}[ht]
\begin{center}
\begin{picture}(400,200)(0,80)
\put(0,287){\makebox(0,0)[lc]{bare scalar}}
\put(0,275){\makebox(0,0)[lc]{$q\bar{q}$ nonets}}
\thinlines
\put(0,260){\line(1,0){400}}
\put(0,140){\line(1,0){400}}
\put(150,260){\makebox(0,0){\Large\bm{\bullet}}}
\put(250,260){\makebox(0,0){\Large\bm{\bullet}}}
\put(350,260){\makebox(0,0){\Large\bm{\bullet}}}
\put(150,240){\makebox(0,0){\bm{\approx 1.4 GeV}}}
\put(250,240){\makebox(0,0){\bm{\approx 1.8 GeV}}}
\put(350,240){\makebox(0,0){\bm{\approx 2.1 GeV}}}
\put(150,275){\makebox(0,0){ground state}}
\put(250,285){\makebox(0,0){first radial}}
\put(250,275){\makebox(0,0){excitation}}
\put(350,285){\makebox(0,0){second radial}}
\put(350,275){\makebox(0,0){excitation}}
\put(203,180){\makebox(0,0){coupling}}
\put(203,170){\makebox(0,0){to meson loops}}
\put(0,165){\makebox(0,0)[lc]{$S$-wave meson-meson}}
\put(0,155){\makebox(0,0)[lc]{resonance nonets}}
\linethickness{1.5pt}
\put(20,140){\makebox(0,0){\Large\bm{\bullet}}}
\put(150,140){\makebox(0,0){\Large\bm{\bullet}}}
\put(250,140){\makebox(0,0){\Large\bm{\bullet}}}
\put(350,140){\makebox(0,0){\Large\bm{\bullet}}}
\put(0,120){\makebox(0,0)[lc]{\bm{\approx 0.8 GeV}}}
\put(85,105){\makebox(0,0)
{\large\bm{\underbrace{
\;\;\;\;\;\;\;\;\;\;\;\;\;\;\;\;\;\;\;\;\;\;\;\;\;\;\;\;}}}}
\put(85,90){\makebox(0,0){pole doubling}}
\end{picture}
\end{center}
\caption[]{Schematic presentation of pole doubling for the light flavours
\cite{BRMDRR86}.
Bare meson nonets turn into nonets of mesonic resonances.
In the process of switching on the model parameter $\lambda$,
one extra nonet of singularities appears in the nonet of $S$-wave
meson-meson scattering amplitudes.}
\label{dubbel}
\end{figure}
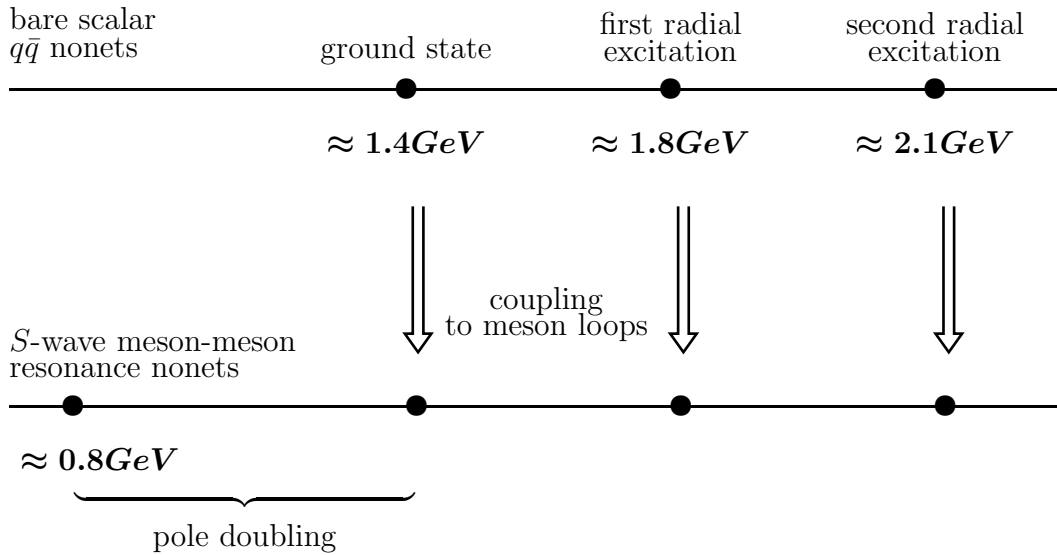

For $S$-wave meson-meson scattering, our model produces, besides
an infinity of $J^{P}=0^{+}$ $S$-matrix poles or resonances,
each stemming from a level in the bare spectrum,
also one additional singularity that is of a completely different nature.
In Fig.~(\ref{dubbel}), we schematically depict this situation.
The nonets of scalar resonances are formed by $P$-wave
$q\bar{q}$ states coupled to pseudoscalar-pseudoscalar and vector-vector
two-meson channels.
There are two isodoublets (non-strange+strange),
which among their decay modes have a $K\pi$ channel. 
Then there is one isotriplet (non-strange+non-strange) involving an
$\eta\pi$ channel.
And finally there is a system of two isosinglet channels, one for $n\bar{n}$
and one for $s\bar{s}$, which mostly couple to each other through their common
$\pi\pi$ and $K\bar{K}$ channels.

Each of the isodoublet states has an extra pole \cite{BRMDRR86} at $727-i263$
MeV, which corresponds to the two doublets of $K_{0}^{\ast}(830)$ resonances.
The three isotriplet states have extra poles  \cite{BRMDRR86}
at $962-i28$ MeV, which describe the triplet of $a_{0}(980)$ resonances.
Finally, the coupled system of isosinglets has extra poles \cite{BRMDRR86}
at $470-i208$ MeV and $994-i20$ MeV, which yield the $f_{0}$(400--1200) and
$f_{0}(980)$ resonances, respectively.

The wave functions of the $K_{0}^{\ast}(830)$, $a_{0}(980)$,
$f_{0}$(400--1200), and $f_{0}(980)$ resonances do not have nodes.
Consequently, in this sense they must be considered {\it ground states}.
The $a_{0}(980)$ and $f_{0}(980)$ resonances are experimentally
well established.
The $f_{0}$(400--1200) resonance is still disputed,
whereas the $K_{0}^{\ast}(830)$ has recently been reported in preliminary
experimental work \cite{Goebel00}.

The nonet of resonances stemming from the nonet of ground states of the bare
meson spectrum around 1.4 GeV consists of the experimentally established
scalars $K_{0}^{\ast}(1430)$, $a_{0}(1450)$, $f_{0}(1370)$, and $f_{0}(1500)$.
The higher nonets are still incomplete.

Scalar mesons, especially the light ones, have been the subject of lively
debate for several decades. We believe that the here described model has all
the necessary ingredients for a reasonable description of the data
(see also Ref.~\cite{KimM99}).

\begin{figure}[ht]
\footnotesize
\begin{center}
\begin{picture}(200,140)(-42.00,-25.20)
\put(25.63,-4.64){\makebox(0,0)[tc]{0.8}}
\put(57.68,-4.64){\makebox(0,0)[tc]{1.0}}
\put(89.72,-4.64){\makebox(0,0)[tc]{1.2}}
\put(-4.64,25.33){\makebox(0,0)[rc]{20}}
\put(-4.64,50.67){\makebox(0,0)[rc]{40}}
\put(-4.64,76.00){\makebox(0,0)[rc]{60}}
\put(157.01,-4.64){\makebox(0,0)[tr]{$GeV$}}
\put(74.49,-16.86){\makebox(0,0)[tc]{invariant $K\pi$ mass}}
\put(-4.64,112.95){\makebox(0,0)[tr]{GeV$^{-2}$}}
\put(4.64,108.94){\makebox(0,0)[tl]{$\sigma$}}
\put(14.42,87.63){\makebox(0,0){$\odot$}}
\put(22.43,85.03){\makebox(0,0){$\odot$}}
\put(32.84,91.69){\makebox(0,0){$\odot$}}
\put(34.45,101.14){\makebox(0,0){$\odot$}}
\put(36.05,86.20){\makebox(0,0){$\odot$}}
\put(37.65,81.96){\makebox(0,0){$\odot$}}
\put(39.25,81.22){\makebox(0,0){$\odot$}}
\put(40.86,78.85){\makebox(0,0){$\odot$}}
\put(42.46,59.87){\makebox(0,0){$\odot$}}
\put(44.06,53.66){\makebox(0,0){$\odot$}}
\put(45.66,61.49){\makebox(0,0){$\odot$}}
\put(47.26,68.97){\makebox(0,0){$\odot$}}
\put(48.87,71.12){\makebox(0,0){$\odot$}}
\put(50.47,78.75){\makebox(0,0){$\odot$}}
\put(54.47,74.72){\makebox(0,0){$\odot$}}
\put(60.88,75.27){\makebox(0,0){$\odot$}}
\put(67.29,72.17){\makebox(0,0){$\odot$}}
\put(73.70,67.08){\makebox(0,0){$\odot$}}
\put(80.11,61.02){\makebox(0,0){$\odot$}}
\put(86.52,59.89){\makebox(0,0){$\odot$}}
\put(92.93,56.57){\makebox(0,0){$\odot$}}
\put(99.33,53.05){\makebox(0,0){$\odot$}}
\put(105.74,52.00){\makebox(0,0){$\odot$}}
\put(111.83,49.71){\makebox(0,0){$\odot$}}
\put(117.60,45.41){\makebox(0,0){$\odot$}}
\put(124.49,36.61){\makebox(0,0){$\odot$}}
\put(131.22,27.74){\makebox(0,0){$\odot$}}
\put(137.79,15.48){\makebox(0,0){$\odot$}}
\put(147.40,0.36){\makebox(0,0){$\odot$}}
\put(30.28,62.29){\makebox(0,0){$\bullet$}}
\put(31.56,79.64){\makebox(0,0){$\bullet$}}
\put(33.65,89.80){\makebox(0,0){$\bullet$}}
\put(34.45,79.21){\makebox(0,0){$\bullet$}}
\put(36.05,97.88){\makebox(0,0){$\bullet$}}
\put(38.13,83.90){\makebox(0,0){$\bullet$}}
\put(40.86,83.52){\makebox(0,0){$\bullet$}}
\put(42.94,62.91){\makebox(0,0){$\bullet$}}
\put(44.38,81.17){\makebox(0,0){$\bullet$}}
\put(46.30,71.11){\makebox(0,0){$\bullet$}}
\put(47.26,77.63){\makebox(0,0){$\bullet$}}
\put(48.87,77.03){\makebox(0,0){$\bullet$}}
\put(52.55,72.49){\makebox(0,0){$\bullet$}}
\put(55.27,70.11){\makebox(0,0){$\bullet$}}
\put(59.28,74.18){\makebox(0,0){$\bullet$}}
\put(63.13,71.90){\makebox(0,0){$\bullet$}}
\put(69.69,66.07){\makebox(0,0){$\bullet$}}
\put(75.94,61.50){\makebox(0,0){$\bullet$}}
\put(82.03,57.68){\makebox(0,0){$\bullet$}}
\put(88.12,56.25){\makebox(0,0){$\bullet$}}
\put(94.53,55.73){\makebox(0,0){$\bullet$}}
\put(100.94,53.68){\makebox(0,0){$\bullet$}}
\put(107.34,52.14){\makebox(0,0){$\bullet$}}
\put(110.55,50.16){\makebox(0,0){$\bullet$}}
\put(113.75,48.28){\makebox(0,0){$\bullet$}}
\put(117.44,45.65){\makebox(0,0){$\bullet$}}
\put(120.48,41.26){\makebox(0,0){$\bullet$}}
\put(123.37,34.86){\makebox(0,0){$\bullet$}}
\put(126.57,25.11){\makebox(0,0){$\bullet$}}
\put(129.78,20.74){\makebox(0,0){$\bullet$}}
\put(133.62,15.87){\makebox(0,0){$\bullet$}}
\put(136.18,9.34){\makebox(0,0){$\bullet$}}
\put(139.87,6.90){\makebox(0,0){$\bullet$}}
\put(143.39,7.13){\makebox(0,0){$\bullet$}}
\put(146.76,3.54){\makebox(0,0){$\bullet$}}
\put(150.12,1.40){\makebox(0,0){$\bullet$}}
\end{picture}
\end{center}
\normalsize
\caption[]{Elastic $S$-wave $K\pi$ cross section.
Data are taken from Refs.~\cite{ECMDLL78,Esta79} ($\odot$) and
\cite{Aston88} ($\bullet$).
The solid line is a model result \cite{BRBW01}.
Maximum of the theoretical cross section is at 832 MeV.}
\label{kappa}
\end{figure}
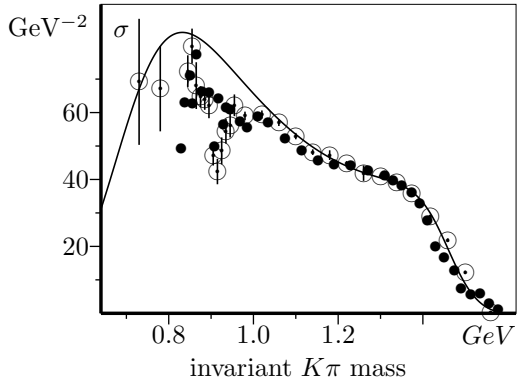
In Fig.~(\ref{kappa}), the prediction of Ref.~\cite{BRBW01} for the $S$-wave
cross section in isodoublet $\pi K$ scattering is shown.
One observes that the central peak lies at 832 MeV.
Nevertheless, the related singularity of the $\pi K$ scattering amplitude we
found at $714-i228$ MeV.  Hence, straightforward relations between bumps in the
cross section and pole positions do not always exist.

Recently it has been recognised that $S$-matrix pole positions have more
consistency than resonant structures in the data.
Values found in the literature for the $\sigma$ and $\kappa$ resonances
are collected and depicted in Table~(\ref{sigkap}).
The world averages for the $f_{0}$(400--1200) and $K^{\ast}_{0}(830)$
pole positions, thereby ignoring the differences in the nature of the
various analyses, are
\begin{displaymath}
f_{0}\;\left[
(506\pm 119)-i(246\pm 85)
\;\;\xrm{MeV}\;\right]\; ;\;\;\;
K^{\ast}_{0}\;\left[
(786\pm 78)-i(255\pm 96)
\;\;\xrm{MeV}\;\right]
\;\;\; .
\end{displaymath}
\begin{table}[ht]
\begin{center}
\begin{tabular}{|rr||c|c|l}
\cline{1-4}\cline{1-4} & & & & \\ [-0.3cm]
date & Ref. & $\sigma$ (MeV) & $\kappa$ (MeV) & \\ [.2cm]
\cline{1-4} & & & & \\ [-0.3cm]
1970 & \cite{BL70} &
 $425-i110$ &  &
\footnotesize
\begin{picture}(200,0)(0,340)
\put(0,202){\makebox(0,0)[bl]{
\begin{picture}(198.43,180.08)(-42.00,0.00)
\put(24.92,150.06){\makebox(0,0)[bc]{0.4}}
\put(66.46,150.06){\makebox(0,0)[bc]{0.6}}
\put(108.00,150.06){\makebox(0,0)[bc]{0.8}}
\put(-4.64,16.50){\makebox(0,0)[rc]{-0.5}}
\put(-4.64,42.29){\makebox(0,0)[rc]{-0.4}}
\put(-4.64,68.07){\makebox(0,0)[rc]{-0.3}}
\put(-4.64,93.85){\makebox(0,0)[rc]{-0.2}}
\put(-4.64,119.64){\makebox(0,0)[rc]{-0.1}}
\put(157.01,150.06){\makebox(0,0)[br]{GeV}}
\put(74.49,162.29){\makebox(0,0)[bc]{$Re(\sqrt{s}\, )$}}
\put(-4.64,4.01){\makebox(0,0)[br]{GeV}}
\put(4.64,8.03){\makebox(0,0)[bl]{$Im(\sqrt{s}\, )$}}
\put(141.31,14.54){\makebox(0,0){\large \bm{\sigma}}}
\put(30.11,117.06){\makebox(0,0){$\bullet$}}
\put(30.74,77.09){\makebox(0,0){$\bullet$}}
\put(41.54,22.95){\makebox(0,0){$\bullet$}}
\put(37.38,58.27){\makebox(0,0){$\bullet$}}
\put(97.61,109.32){\makebox(0,0){$\bullet$}}
\put(39.46,91.79){\makebox(0,0){$\circ$}}
\put(130.84,55.18){\makebox(0,0){$\bullet$}}
\put(29.08,50.02){\makebox(0,0){$\bullet$}}
\put(18.69,51.31){\makebox(0,0){$\bullet$}}
\put(46.94,81.74){\makebox(0,0){$\bullet$}}
\put(22.22,66.78){\makebox(0,0){$\bullet$}}
\put(39.46,80.96){\makebox(0,0){$\bullet$}}
\put(57.94,97.72){\makebox(0,0){$\bullet$}}
\put(39.25,99.27){\makebox(0,0){$\bullet$}}
\put(66.88,94.89){\makebox(0,0){$\bullet$}}
\put(33.65,86.89){\makebox(0,0){$\bullet$}}
\put(29.49,90.50){\makebox(0,0){$\bullet$}}
\put(34.27,84.83){\makebox(0,0){$\bullet$}}
\put(29.08,63.69){\makebox(0,0){$\bullet$}}
\put(49.43,78.13){\makebox(0,0){$\bullet$}}
\put(34.27,88.44){\makebox(0,0){$\bullet$}}
\put(45.28,62.66){\makebox(0,0){$\bullet$}}
\put(93.87,113.71){\makebox(0,0){$\bullet$}}
\put(41.12,103.65){\makebox(0,0){$\bullet$}}
\put(35.10,83.28){\makebox(0,0){$\bullet$}}
\put(63.34,95.66){\makebox(0,0){$\bullet$}}
\put(52.96,105.46){\makebox(0,0){$\bullet$}}
\put(47.04,81.95){\makebox(0,0){$\cdot$}}
\end{picture}
}}
\put(0,10){\makebox(0,0)[bl]{
\begin{picture}(198.43,180.08)(-42.00,0.00)
\put(28.97,150.06){\makebox(0,0)[bc]{0.7}}
\put(77.25,150.06){\makebox(0,0)[bc]{0.8}}
\put(-4.64,30.01){\makebox(0,0)[rc]{-0.4}}
\put(-4.64,58.86){\makebox(0,0)[rc]{-0.3}}
\put(-4.64,87.71){\makebox(0,0)[rc]{-0.2}}
\put(-4.64,116.57){\makebox(0,0)[rc]{-0.1}}
\put(157.01,150.06){\makebox(0,0)[br]{GeV}}
\put(74.49,162.29){\makebox(0,0)[bc]{$Re(\sqrt{s}\, )$}}
\put(-4.64,4.01){\makebox(0,0)[br]{GeV}}
\put(4.64,8.03){\makebox(0,0)[bl]{$Im(\sqrt{s}\, )$}}
\put(141.31,14.54){\makebox(0,0){\large \bm{\kappa}}}
\put(12.07,24.24){\makebox(0,0){$\bullet$}}
\put(77.25,133.88){\makebox(0,0){$\bullet$}}
\put(42.01,69.54){\makebox(0,0){$\circ$}}
\put(113.46,48.76){\makebox(0,0){$\bullet$}}
\put(62.77,73.29){\makebox(0,0){$\bullet$}}
\put(67.11,50.21){\makebox(0,0){$\bullet$}}
\put(130.84,99.83){\makebox(0,0){$\bullet$}}
\put(75.80,86.27){\makebox(0,0){$\bullet$}}
\put(32.83,57.42){\makebox(0,0){$\bullet$}}
\put(127.95,66.65){\makebox(0,0){$\bullet$}}
\put(35.73,79.64){\makebox(0,0){$\bullet$}}
\put(70.71,71.79){\makebox(0,0){$\cdot$}}
\end{picture}
}}
\end{picture}
\normalsize\\
1971 & \cite{BZJ71} &
 $428-i265$ & & \\
1972 & \cite{BFP72} &
 $480-i475$ & & \\
1973 & \cite{IZJZ73} &
 $460-i338$ & $665-i420$ & \\
1982 & \cite{Scadron82} &
 $750-i140$ & $800-i 40$ & \\
1986 & \cite{BRMDRR86} &
 $470-i208$ & $727-i263$ & \\
1987 & \cite{AMP87} &
 $910-i350$ & & \\
1994 & \cite{AS94} &
 $420-i370$ & & \\
1994 & \cite{ZB94} &
 $370-i365$ & & \\
1994 & \cite{KLM94} &
 $506-i247$ & & \\
1995 & \cite{JPHS95} &
 $387-i305$ & & \\
1996 & \cite{TR96} &
 $470-i250$ & absent & \\
1996 & \cite{HSS96} &
 $559-i185$ & & \\
1997 & \cite{OO97} &
 $469-i179$ & & \\
1997 & \cite{IIITT97} &
 $602-i196$ & $875-i335$ & \\
1998 & \cite{OOP98} &
 $442-i227$ & $770-i250$ & \\
1998 & \cite{LMZ98} &
 $422-i213$ & & \\
1999 & \cite{Han99} &
 $445-i235$ & & \\
1999 & \cite{AN99} &
 $420-i317$ & & \\
1999 & \cite{KLL99} &
 $518-i261$ & & \\
1999 & \cite{OO99} &
 $445-i221$ & $779-i330$ & \\
1999 & \cite{BFSS99} &
 & $911-i158$ & \\
2000 & \cite{XZ00} &
 $498-i321$ & & \\
2000 & \cite{KLR00} &
 $732-i123$ & & \\
2000 & \cite{Goebel00} &
 $478-i162$ & $797-i205$ & \\
2000 & \cite{JOP00} &
 & $708-i305$ & \\
2001 & \cite{AXZS01} &
 $449-i241$ & & \\
2001 & \cite{Takam01} &
 $585-i193$ & $905-i273$ & \\
2001 & \cite{KII01} &
 $535-i155$ & & \\
2001 & \cite{BRBW01} &
 & $714-i228$ & \\
\cline{1-4}\cline{1-4}
\end{tabular}
\end{center}
\caption[]{
Results and references for $\sigma$ and $\kappa$
poles in the complex $\sqrt{s}$ plane for isosinglet $\pi\pi$ and
isodoublet $\pi K$ $S$-wave scattering, respectively.
The upper figure shows the locations of the $\sigma$ poles,
the lower one for the $\kappa$ poles.
World averages are indicated in the figures.
Our model predictions \cite{BRMDRR86} are indicated by $\circ$.}
\label{sigkap}
\end{table}
\clearpage

\section{Radiative transitions}

Multichannel wave functions are neither very suitable for figures, nor for
comprehensive tables. Radiative decays, however, represent a rather sensitive
test for their correctness. Such tests has been carried out in
Ref.~\cite{VDB91} for the $c\bar{c}$ and $b\bar{b}$ sectors of the model.
It involves the development of a formalism for radiative transitions of
multicomponent systems with quarkonium and meson-meson channels.
In Tables~(\ref{ccrt}) and (\ref{bbrt}), we compare
the theoretical predictions of Ref.~\cite{VDB91} for 
the $c\bar{c}$ and $b\bar{b}$ systems, respectively, with the presently
available data.
\begin{table}[ht]
\begin{center}
\begin{tabular}{|c|c||r|r|}
\hline\hline
state & decay product & model \cite{VDB91} & experiment \cite{PDG}\\
\hline & & & \\ [-0.3cm]
 &  & keV & keV \\ [.2cm]
$J/\psi (1S)$ & $\gamma\eta_{c}(1S)$ & 1.67 & 1.13$\pm$0.41\\
\hline & & & \\ [-0.3cm]
$\psi (2S)$ & $\gamma\chie_{c0}(1P)$ & 31.7 & 25.8$\pm$5.4 \\
            & $\gamma\chie_{c1}(1P)$ & 49.8 & 24.1$\pm$4.9 \\
            & $\gamma\chie_{c2}(1P)$ & 38.5 & 21.6$\pm$4.6 \\
            & $\gamma\eta_{c}(1S)$  & 1.41 & 0.78$\pm$0.25 \\
\hline & & & \\ [-0.3cm]
$\chie_{c0}(1P)$ & $\gamma J/\psi (1S)$  & 52 & 98$\pm$43 \\
\hline & & & \\ [-0.3cm]
$\chie_{c1}(1P)$ & $\gamma J/\psi (1S)$  & 210 & 240$\pm$52 \\
\hline & & & \\ [-0.3cm]
$\chie_{c2}(1P)$ & $\gamma J/\psi (1S)$  & 228 & 270$\pm$46 \\
\hline\hline
\end{tabular}
\end{center}
\caption[]{Comparison of the model results for $c\bar{c}$
radiative transitions to experiment.}
\label{ccrt}
\end{table}
\begin{table}[ht]
\begin{center}
\begin{tabular}{|c|c||r|r|}
\hline\hline
state & decay product & model \cite{VDB91} & experiment \cite{PDG}\\
\hline & & & \\ [-0.3cm]
 &  & keV & keV \\ [.2cm]
$\Upsilon (2S)$ & $\gamma\chie_{b0}(1P)$ & 1.6 & 1.7$\pm$0.5 \\
            & $\gamma\chie_{b1}(1P)$ & 2.1 & 3.0$\pm$0.8 \\
            & $\gamma\chie_{b2}(1P)$ & 2.6 & 3.1$\pm$0.8 \\
\hline & & & \\ [-0.3cm]
$\Upsilon (3S)$ & $\gamma\chie_{b0}(2P)$ & 1.6 & 1.4$\pm$0.3 \\
                & $\gamma\chie_{b1}(2P)$ & 2.6 & 3.0$\pm$0.6 \\
                & $\gamma\chie_{b2}(2P)$ & 3.1 & 3.0$\pm$0.6 \\
\hline\hline
\end{tabular}
\end{center}
\caption[]{Comparison of the model results for $b\bar{b}$
radiative transitions to experiment.}
\label{bbrt}
\end{table}

The agreement of the model's predictions with the data is fairly good
and has much improved since their publication in 1991, which is an additional
argument in favour of the usefulness of the employed formalism.
\clearpage

\section{Data analysis}

The here described model produces non-exotic bare mesons with all possible
angular quantum numbers and an infinity of radial excitations, which,
through the universal parameters $\lambda$ and $a$,
turn into the experimental spectra of all existing mesonic bound states and
resonances.
However, in spite of the model's success to reproduce
a host of experimental data with a very limited number of parameters, it is
clearly not suited as a tool for data analysis, owing to the specific model
choice of the confining $q\bar{q}$ potential, and furthermore the rather
complicated matrix expressions needed to obtain $S$-matrix-related observables.

In recent work \cite{BRBW01}, we have indicated how the model's formalism may
be applied to data analysis.
The bare meson masses and their couplings to the meson-meson channels
enter as free fit parameters in this approach.
In particular, we have shown that the $S$-matrix singularities
describing the light scalar mesons also appear here.
Hence, this simplified model, which can be solved analytically, serves well as
a laboratory for studying the general properties of meson-meson scattering.
Moreover, for a more detailed description of specific processes,
parts of the analytic solutions can be substituted by adjustable parameters.

\section{Hybrids}

In a recent work \cite{Close01}, Frank Close suggests the use
of a model similar to ours for the study of hybrids.
Certainly, as the objects which we here call constituent quarks can exchange
quark and colour degrees of freedom, one may naturally think of different quark
and/or colour configurations. In this sense, also exotic mesonic resonances
could, in principle, be accommodated.
\clearpage

\section{Conclusions}

Bound states and resonances are most conveniently characterised by
singularities of the scattering matrix as a function of the total invariant
mass. It has the advantage that even in cases where the resonating behaviour
of the scattering cross sections is not very clear from experiment,
their existence clarifies the classification of well-established
resonances into flavour multiplets \cite{BR01b}.
However, a disadvantage is that each different theoretical model locates
such poles at different complex values.
Hence, no agreement can be expected on their positions.
Nevertheless, when several analyses, describing the same phenomenon, obtain
poles at more or less the same values, like those given in
Table~(\ref{sigkap}), then no doubt can exist about their existence in the true
model of Nature.

Knowledge about how the poles depend on the parameters of strong interactions
in each model could, moreover, lessen the discrepancies.
Since it is indispensable to gain experience on the behaviour of
mesonic $S$-matrix singularities, it is useful to dispose of analytical
expressions that incorporate most of our knowledge on strong interactions,
and are also capable of describing the data to a moderate degree of accuracy.
The outlined simplified model serves well for that purpose.
\vspace{0.3cm}

{\bf Acknowledgement}:
This work has been partly supported by the
{\it Fun\-da\-\c{c}\~{a}o para a Ci\^{e}ncia e a Tecnologia}
\/of the {\it Minist\'{e}rio da
Ci\^{e}ncia e da Tecnologia} \/of Portugal,
under contract numbers
POCTI/\-35304/\-FIS/\-2000,
and
CERN/\-P/\-FIS/\-40119/\-2000.

\end{document}